\documentclass[aps,prl,twocolumn,amsmath,amssymb,superscriptaddress]{revtex4-1}

\usepackage{graphicx,tabularx}
\usepackage{multirow}
\usepackage{dcolumn}
\usepackage{amssymb,amscd,xypic,bm,wasysym}
\usepackage{float}
\usepackage{cleveref}

\makeatletter
\newsavebox{\@brx}
\newcommand{\llangle}[1][]{\savebox{\@brx}{\(\m@th{#1\langle}\)}%
	\mathopen{\copy\@brx\mkern2mu\kern-0.9\wd\@brx\usebox{\@brx}}}
\newcommand{\rrangle}[1][]{\savebox{\@brx}{\(\m@th{#1\rangle}\)}%
	\mathclose{\copy\@brx\mkern2mu\kern-0.9\wd\@brx\usebox{\@brx}}}
\makeatother

\begin{document}

\title{Spin Hall effects without spin currents in magnetic insulators}

\author{Hua Chen}
    \affiliation{Department of Physics, Colorado State University, Fort Collins, CO 80523, USA}
    \affiliation{School of Advanced Materials Discovery, Colorado State University, Fort Collins, CO 80523, USA}
\author{Qian Niu}
    \affiliation{Department of Physics, University of Texas at Austin, Austin, TX 78712, USA}
\author{Allan H. MacDonald}
    \affiliation{Department of Physics, University of Texas at Austin, Austin, TX 78712, USA}
    
\begin{abstract}
The spin Hall effect (SHE) is normally discussed in terms of a spin current,
which is ill-defined in strongly spin-orbit-coupled systems because of spin non-conservation. 
In this work we propose an alternative view of SHE phenomena by relating them to 
a spin analog of charge polarization induced by an electric field.    
The spin density polarization is most conveniently defined in insulators,
which can have a SHE if they break time-reversal symmetry, {\it i.e.} if they are magnetic. The reciprocal of this SHE is a counterpart of the inverse SHE (ISHE), and is manifested in magnetic insulators as a charge polarization induced by a Zeeman field gradient. We use a modified Kane-Mele model to illustrate the  
magnetic spin Hall effect, and to discuss its bulk-boundary relationship.
\end{abstract}

\maketitle

The spin Hall effect (SHE) in a nonmagnetic conductor is normally defined as transverse spin-current response to an electric field \cite{dyakonov_1971,hirsch_1999,kato_2004,wunderlich_2005}. It was originally proposed \cite{dyakonov_1971,hirsch_1999} as a spin-analog of the charge-transport anomalous Hall effect (AHE) \cite{nagaosa_2010} that survives in non-magnetic systems, and described as a consequence of spin-dependent (skew) disorder scattering. When spin-orbit coupling in the band Hamiltonian is weak enough that the concept of a spin-dependent local chemical potential is valid, the notion of a spin current is well defined, even though spin is not strictly conserved. However, recent advances in SHE-related spintronics have consistently identified materials
with strong intrinsic spin-orbit coupling as favorable for potential applications. The notion of a spin current is generally speaking not useful when the typical Bloch state spin splitting produced by spin-orbit coupling is larger than $\hbar/\tau$, where $\tau$ is the disorder scattering time. In this case it is not normally possible to derive modified continuity equations for spin, certainly not ones which simply add a decay term to account for spin non-conservation \cite{shi_2006,chen_2017}, and the introduction of phenomenological quantities like an interface spin-mixing conductance to explain experiments may not be well justified. This Letter aims at providing an appropriate descriptor of the spin Hall effect in the materials with strong intrinsic spin-orbit coupling that are of greatest interest in modern spintronics.    

The SHE continues to be a well-defined observable effect, however, if it is identified with spin response at sample boundaries or 
with an influence on the collective magnetization of an adjacent magnetic system, {\it i.e.} with the physical quantities that are in any event ultimately measured \cite{saitoh_2006, kimura_2007, liu_2012, jungwirth_2012}. In contrast to the charge Hall effect case, long-range electromagnetic interactions between local spin densities at opposite boundaries do not play an essential role in establishing the spin-Hall steady state. It is this fact that makes it possible to describe the SHE as a local linear response of spin density at the boundaries of a conductor to electric field, without involving the concept of spin currents \cite{freimuth_2014,hals_2015,kimata_2018}. The disadvantage of such a description, however, is that one has to explicitly consider the boundary and cannot easily identify bulk quantities to explain the effect. When the notion of a spin current (density) is applicable, its use is attractive because it is an intensive quantity that stays finite in the thermodynamic limit, where complicating influences from the boundary approach zero. 

Motivated by these considerations, we propose use of the \emph{polarization} of spin density, 
\begin{equation}
\overleftrightarrow{p}_s =  \int d{\bf r} \, \frac{{\bf s}\otimes {\bf r}}{V}
\end{equation}
where $V$ is the volume of the system, as a bulk descriptor of the SHE. This quantity is finite in the thermodynamic limit when there is a net area density of opposite spin at opposite boundaries. We note that one of us has previously used $\overleftrightarrow{p}_s$ to propose an alternative definition of the spin current \cite{shi_2006}. Obviously spin non-conservation does not impact the definition of $\overleftrightarrow{p}_s$. The remainder of this Letter will focus on understanding the properties of spin density polarization in the ground state of a periodic system and its response to electric fields, which corresponds to the SHE. In particular, we will identify a counterpart of the SHE which can occur in magnetic insulators in the absence of transport currents, therefore corresponding to an infinite spin Hall angle in the spin current language. We are also aware of related literatures that discuss similar constructions of the spatial distribution of spin density \cite{spaldin_2013, gao_2017}, but possible connections with the SHE have not been discussed.

In this work we only consider spin density polarization in insulators, where it can be conveniently described using exponentially localized ground state Wannier functions. Possible generalization to metallic systems is mentioned at the end of the Letter. The formulation presented below has much in common with the modern theory of electric polarization \cite{king-smith_1993}. In a finite insulating system the spin density polarization
\begin{eqnarray}
\langle p_s \rangle_{\alpha\beta} = \frac{1}{V} \sum_{n{\bf R}} \langle n{\bf R} |s_\alpha r_\beta |n{\bf R}\rangle,
\end{eqnarray}
where $\alpha,\beta$ label $x,y,z$, and $n$ and unit cell position $\bf R$ are Wannier orbital labels. Ignoring for now any potential difference between localized orbitals near the boundary and Wannier orbitals deep inside the bulk (see below), we have
\begin{eqnarray}\label{eq:psreal}
\langle p_s \rangle_{\alpha\beta} = \frac{1}{V} \sum_{n{\bf R}} \langle n{\bf R} |s_\alpha (r_\beta - R_\beta) |n{\bf R}\rangle \\\nonumber
+ \frac{1}{V} \sum_{n{\bf R}} \langle n{\bf R} |s_\alpha |n{\bf R}\rangle R_\beta,
\end{eqnarray}
where the first term respects translational symmetry and can be Fourier-transformed to a momentum space expression, while the 2nd term is ill-behaved if the total spin density in each unit cell is nonzero. We ignore this contribution below since it vanishes except in ferromagnets and ferrimagnets, and can always be eliminated by a judicious choice of origin. The momentum space expression for the ground state $\overleftrightarrow{p}_s$ is therefore
\begin{eqnarray}\label{eq:pskspace}
\langle p_s \rangle_{\alpha\beta} =\int [d{\bf k}] \sum_{n} i \langle u_{n\bf k} |s_\alpha|\partial_{k_\beta} u_{n\bf k}\rangle, 
\end{eqnarray}
where $|u_{n\bf k}\rangle$ is the periodic part of the Bloch state for the $n$th occupied Wannier orbital. 

Several comments on differences between Eq.~\ref{eq:pskspace} and the corresponding ground state electric polarization expression \cite{king-smith_1993} are in order:

\noindent (i)  The ground state spin density polarization is not gauge invariant, and does not have a gauge invariant ``quantum" in general. It is well known that the electric polarization is not gauge invariant, but it can only change in integer multiples of a gauge-invariant quantum, which corresponds to moving a single electron by one unit lattice vector \cite{king-smith_1993}. We can see that he spin density polarization does not have this property in general by multiplying each $|\partial_{k_\beta} u_{n\bf k}\rangle$ by a phase factor $e^{i\phi_{n\bf k}}$, where $\phi_{n\bf k}$ is periodic (mod $2\pi$) in the Brillouin zone. $\langle \overleftrightarrow{p}_s \rangle$ then changes by 
\begin{eqnarray}
\Delta\langle p_s \rangle_{\alpha\beta} = - \int [d{\bf k}] \sum_{n} \langle u_{n\bf k} |s_\alpha| u_{n\bf k}\rangle \partial_{k_\beta} \phi_{n\bf k},
\end{eqnarray}
which is not quantized because of the momentum dependence of the spin expectation value in Bloch states when spin-orbit coupling is not negligible.

\noindent (ii) The ground state spin density polarization can be separated into gauge-invariant and gauge-dependent contributions by inserting $1 = P_{\bf k} + Q_{\bf k}$, where $P_{\bf k} = \sum_n |u_{n\bf k} \rangle \langle u_{n\bf k} |$ and $Q_{\bf k} = 1-P_{\bf k}$, between the spin operator and the $k$-derivative in Eq.~\ref{eq:pskspace} to obtain
\begin{eqnarray}
\label{eq:separation}
\langle p_s \rangle_{\alpha\beta} = -\int[d{\bf k}] {\rm Im Tr}[P_{\bf k} s_\alpha (\partial_{k_\beta} P_{\bf k})] \\\nonumber
+ \int[d{\bf k}] \sum_{nm} (s_\alpha)_{nm} (A_\beta)_{mn} \\\nonumber
\equiv (\langle p_s \rangle_{gi})_{\alpha\beta} + (\langle p_s \rangle_{gd})_{\alpha\beta}.
\end{eqnarray}
The first term in Eq.~\ref{eq:separation} is invariant under any $k$-dependent unitary transformations within the occupied space, since it is a trace of the product of gauge-invariant operators \cite{ceresoli_2006}, whereas the 2nd term is explicitly gauge dependent since it involves matrix elements of the non-Abelian Berry connection ${\bf A}_{mn} \equiv i\langle u_{m\bf k} | \nabla | u_{n \bf k}\rangle$ in the occupied space. Since $\partial_{k_\beta} P_{\bf k}$ contains only cross-gap matrix elements, $\langle \overleftrightarrow{p}_s \rangle_{gi}$ is non-zero only in the presence spin-orbit coupling. The electric polarization does not have a counterpart of $\langle \overleftrightarrow{p}_s \rangle_{gi}$ since it does not involve cross-gap matrix elements of position. From the symmetry point of view, a non-zero $\langle \overleftrightarrow{p}_s \rangle_{gi}$ requires spatial inversion symmetry breaking, while it is not necessary for $\langle \overleftrightarrow{p}_s \rangle_{gd}$ (see below).

\noindent (iii) The second term in Eq.~\ref{eq:psreal} can provide a boundary contribution to $\langle \overleftrightarrow{p}_s \rangle$ if the localized wavefunctions at the boundary have a different total spin density per unit cell than the bulk Wannier orbitals. This is similar to the case of electric polarization which can be changed by moving an electron from one surface of a finite system to the opposite surface. In general there are no definite connections between bulk and boundary contributions for $\langle \overleftrightarrow{p}_s \rangle$. We nevertheless note that boundary plays a ``gauge fixing" role since there is no gauge problem for finite systems.

The simplest example of a ground state with non-zero spin density polarization is a G-type N\'{e}el antiferromagnet. When spin-orbit coupling and quantum fluctuations are ignored, the spin component along the direction of the N\'{e}el vector is a good quantum number, and $\langle \overleftrightarrow{p}_s \rangle$ modulo a quantum is invariant under unitary transformations in the occupied space that do not mix spin-up and spin-down. For a single-band nearest-neighbor hopping model with staggered exchange fields on the fcc sublattices of a simple-cubic lattice (which does have spatial inversion symmetry), we can evaluate $\langle \overleftrightarrow{p}_s \rangle$ by choosing a smooth gauge that does not mix up and down spins \cite{supp}. In agreement with expectations based on summing classical spins we find that $\langle p_s \rangle_{zx} = \frac{\hbar a}{2 V_m}$ \cite{supp}, where $a$ is the nearest neighbor distance, $V_m=2a^3$ is the volume of the magnetic unit cell, and the bond between the two sites in the unit cell is chosen to be along $\hat{x}$. Physically this result corresponds to the observation that a finite slab perpendicular to {\it e.g.} $\hat{n}=[111]$, obtained by repeating the magnetic unit cell, has a nonzero area spin density equal to $\frac{\hbar a \hat{n}\cdot \hat{x}}{2 V_m}$. 

We next discuss the linear response of the spin density polarization to external electric fields, which is relevant to the SHE. We adopt a quantum kinetic formalism \cite{culcer_2017} that in the insulating case reduces to the standard density matrix perturbation theory \cite{mcweeny_1962}. In the steady state the perturbed density matrix must satisfy
\begin{eqnarray}
[H_0,\rho^{(1)}] + [H_e,\rho^{(0)}] = 0,
\end{eqnarray}
where $H_0, \rho^{(0)}$ are the Hamiltonian and the density matrix in the absence of perturbation and $\rho^{(1)}$ is the change of the density matrix due to a static electric field perturbation $H_e$. In the $H_0$ eigenstate basis $|m\rangle$, 
\begin{eqnarray}
\rho^{(1)}_{mn} = \left[ \rho_n^{(0)} - \rho_m^{(0)} \right] \frac{(H_e)_{mn}}{\epsilon_n - \epsilon_m}.
\end{eqnarray}
For an insulator $\rho^{(1)}$ has only cross-gap matrix elements \cite{mcweeny_1962}. To consider a homogeneous electric field, we take the $\bf q\rightarrow 0$ limit of the following scalar potential
\begin{eqnarray}
\phi({\bf r}) = -\frac{E}{q} \sin({\bf q}\cdot {\bf r}),
\end{eqnarray} 
where $\bf q$ is parallel to $\bf E$. The perturbed spin density polarization at $\bf q$ is the Fourier transform of 
\begin{eqnarray}
\delta\langle \overleftrightarrow{p}_s\rangle ({\bf r}) = \langle {\bf r}| {\rm Tr}[\rho^{(1)}{\bf s}\otimes({\bf r}-{\bf R})]|{\bf r}\rangle,
\end{eqnarray}
where the trace is taken over all band indices. By taking the ${\bf q}\rightarrow 0$ limit we finally obtain
\begin{eqnarray}\label{eq:dps}
&&\delta\langle \overleftrightarrow{p}_s\rangle = \\\nonumber
&&e{\bf E}\cdot \Bigl [\sum_{\substack{n\in o\\m\in u\\l\in u}}\frac{{\rm Re}({\bf A}_{nm} {\bf s}_{ml}{\bf A}_{ln})}{\epsilon_{m{\bf k}}-\epsilon_{n{\bf k}}} + \sum_{\substack{n\in o \\ m\in u \\ l\in o}} \frac{{\rm Re}({\bf A}_{mn} {\bf s}_{nl}{\bf A}_{lm})}{\epsilon_{m{\bf k}}-\epsilon_{n{\bf k}}} \Bigr ]\\\nonumber
&&+ e{\bf E}\cdot \Bigl [\sum_{\substack{n\in o \\ m\in u \\ l\in o}} \frac{{\rm Re}({\bf A}_{nm} {\bf s}_{ml}{\bf A}_{ln})}{\epsilon_{m{\bf k}}-\epsilon_{n{\bf k}}} + \sum_{\substack{n\in o \\ m\in u \\ l\in u}} \frac{{\rm Re}({\bf A}_{mn} {\bf s}_{nl}{\bf A}_{lm})}{\epsilon_{m{\bf k}}-\epsilon_{n{\bf k}}} \Bigr ]\\\nonumber
&&\equiv \delta\langle \overleftrightarrow{p}_s \rangle_{gi} + \delta\langle \overleftrightarrow{p}_s \rangle_{gd},
\end{eqnarray}
where $o, u$ stand for occupied and unoccupied states respectively, and the scalar product is between $\bf E$ and the first $\bf A$ in each term. The subscripts $gi$ and $gd$ identify the gauge invariant and gauge dependent contributions. Several comments on this result are in order:

\noindent(i) $\delta\langle \overleftrightarrow{p}_s \rangle_{gi}$ and $\delta\langle \overleftrightarrow{p}_s \rangle_{gi}$ are again obtained by inserting $1=P_{{\bf k}}+Q_{{\bf k}}$ between $\bf s$ and $\bf r$. The gauge invariance of $\delta\langle \overleftrightarrow{p}_s \rangle_{gi}$ can be established by recognizing that $\rho^{(1)}_{\bf k}$ transforms as $U_{\bf k}^\dag \rho^{(1)}_{\bf k} U_{\bf k}$ under a unitary transformation $U_{\bf k}$ in the occupied space \cite{supp}.

\noindent(ii) If we consider an isolated band $n$, and neither ${\bf s}_{nn}({\bf k})$ nor ${\bf A}_{nn}({\bf k})$ changes rapidly with $\bf k$, $\delta\langle \overleftrightarrow{p}_s \rangle_{gi}$ can be understood as a contribution in which the electric field changes the spin density polarization by shifting the Wannier orbital of band $n$. In contrast, $\delta\langle \overleftrightarrow{p}_s \rangle_{gd}$ is due to the change of the total spin of band $n$ by the electric field, and thus requires spin-orbit coupling. Only $\delta\langle \overleftrightarrow{p}_s \rangle_{gd}$ involves cross-gap matrix elements of $\bf s$. 

\noindent(iii) Obviously time-reversal symmetry must be broken for $\delta\langle \overleftrightarrow{p}_s \rangle$ to be nonzero. Spatial inversion symmetry breaking is however not necessary because of $\bf E$. For a given crystal lattice, those components of $\delta\langle \overleftrightarrow{p}_s \rangle$ that vanish can be identified using symmetry considerations \cite{birss_book}.  Non-zero components of $\delta\langle \overleftrightarrow{p}_s \rangle$ with spatial Cartesian directions perpendicular to the electric field can be identified with the SHE, which we call in the insulating case magnetic SHE since the system must be magnetic.

\noindent(iv) There is a nontrivial boundary contribution to $\delta\langle \overleftrightarrow{p}_s \rangle$ even in the thermodynamic limit:
\begin{eqnarray}
\delta\langle \overleftrightarrow{p}_s \rangle_b = \frac{1}{V}\sum_{n\bf R} \langle n{\bf R} | \rho^{(1)}_b {\bf s} \otimes ({\bf r} - {\bf R}) |n{\bf R}\rangle \\\nonumber
+ \frac{1}{V}\sum_{n\bf R} \langle n{\bf R} | \rho^{(1)}_b {\bf s} |n{\bf R}\rangle\otimes {\bf R},
\end{eqnarray}
where the subscript $b$ distinguishes sites near the boundary. Although the 1st term vanishes in the thermodynamic limit, the 2nd term does not if the electric field can induce nonzero total spin densities in each unit cell near the boundary. Note this contribution involves only diagonal matrix elements of $\bf r$, and therefore corresponds to $\delta\langle \overleftrightarrow{p}_s \rangle_{gd}$. Because of spin non-conservation this boundary contribution cannot in general be
expressed in terms of bulk quantities. We emphasize again that gauge invariance is always preserved in finite systems. The origin of the gauge dependence in the bulk expressions is that the step of replacing $\bf r$ by ${\rm i}d/d{\bf k}$ is not always justified in the thermodynamic limit. 

To give a concrete example of the physics discussed above, we consider the insulating state of a modified version of the Kane-Mele model \cite{kane_2005, kane_2005_2}. Because $s_z$ is conserved in the standard Kane-Mele model, the spin Hall conductivity for the $s_z$ spin current is well defined. The model does not have a finite spin density polarization induced by an electric field, however, since it is time-reversal invariant. In fact, a finite transverse spin current leads to a constant $\dot{s}_z$ at the transverse boundary, so the system has no steady state in the presence of a longitudinal field. We therefore add a small in-plane Zeeman field to break $s_z$ conservation as well as time-reversal symmetry. The full tight-binding Hamiltonian on a honeycomb lattice is (continuous version of the model is discussed in \cite{supp})
\begin{eqnarray}\nonumber
&&H = \sum_{\langle i\alpha,j\beta \rangle\gamma} t c_{i\alpha\gamma}^\dag c_{j\beta\gamma} + \sum_{\llangle  i\alpha,j\alpha \rrangle\gamma\delta } {\rm i}t_2 \nu_{ij}\sigma^z_{\gamma\delta} c^\dag_{i\alpha\gamma} c_{j\alpha\delta}\\    
&&+ \sum_{i\alpha\gamma\delta} J \sigma^x_{\gamma\delta} c^\dag_{i\alpha\gamma} c_{i\alpha\delta},
\end{eqnarray}
where $i,j$ label unit cell, $\alpha,\beta$ label sublattices, and $\gamma,\delta$ label spin. $\langle \rangle$ and $\llangle\rrangle$ mean nearest and next nearest neighbors, respectively. $\nu_{ij} = {\rm sgn}[({\bf r}_{im}\times{\bf r}_{mj})\cdot\hat{z}]$, where $m$ is the common nearest neighbor of site $i,j$ (the sublattice index is left implicit). $J$ is the strength of the in-plane Zeeman field along $\hat{x}$. The primitive lattice vectors are chosen to be ${\bf a}_1 = \sqrt{3} \hat{x}$, ${\bf a}_2 = \frac{\sqrt{3}}{2}\hat{x} + \frac{3}{2}\hat{y}$, with sublattices $A$ and $B$ located at the origin and $\frac{\sqrt{3}}{2}\hat{x} + \frac{1}{2}\hat{y}$, respectively. The nearest neighbor bond length is set to 1.

To calculate $\delta\langle \overleftrightarrow{p}_s \rangle$ without gauge ambiguities we consider a zigzag ribbon with finite extent in the transverse ($\hat{y}$) direction and evaluate the spin polarization response to a longitudinal electric field. In the tight-binding basis the position operator $y$ is a diagonal matrix in a Wannier representation, with each non-zero matrix element given by the $y$ coordinate of the corresponding site in the ribbon unit cell. (Note that off-diagonal matrix elements of the position operator in the Wannier basis are present in tight-binding models obtained microscopically using modern Wannier techniques.) 

\begin{figure}[h]
	\begin{center}
		\includegraphics[width=2.6 in]{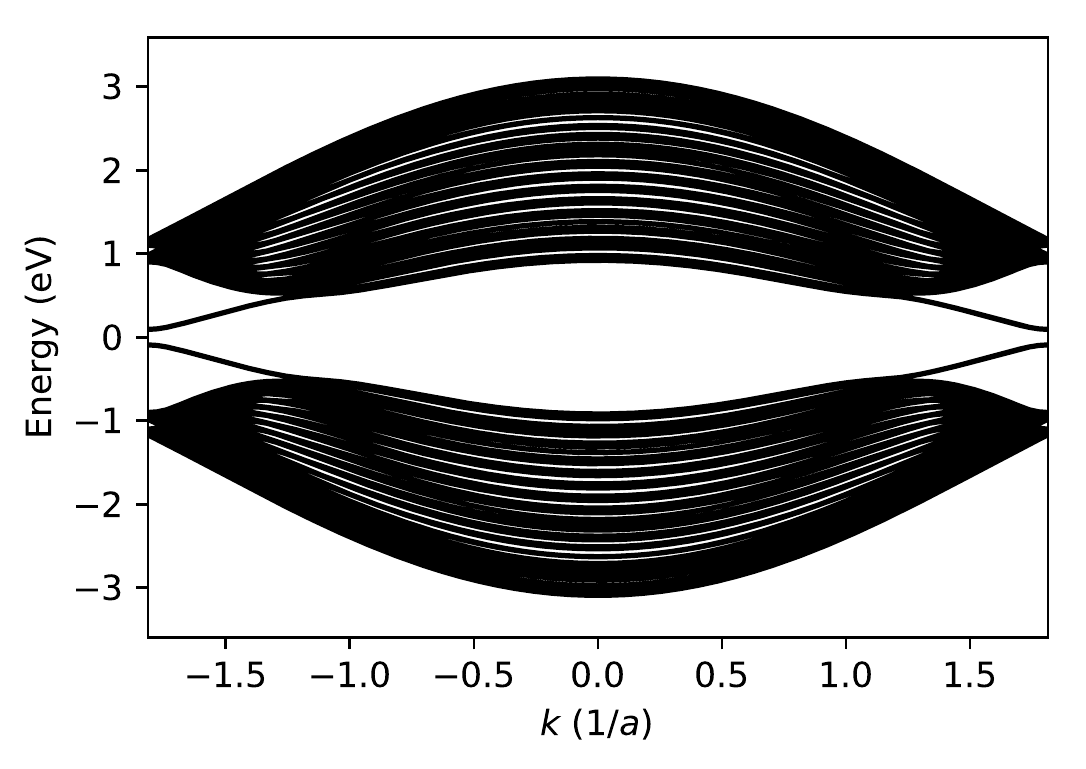}
	\end{center}
	\caption{Band structure for a zigzag ribbon with $N_y=20$, $t=1$, $t_2=0.1$, $J = 0.1$.}
	\label{fig:KMribbon_band}
\end{figure}

Figure~\ref{fig:KMribbon_band} plots the band structure of a ribbon with $N_y=20$ unit cells in the $\hat{y}$ direction. A small gap is opened at the 1D Brillouin zone boundary by the in-plane Zeeman field, which enables us to calculate $\delta\langle \overleftrightarrow{p}_s \rangle$ using Eq.~\ref{eq:dps}. Note that the first ${\bf A}$ in each term of Eq.~\ref{eq:dps} can be expressed as $i\hbar{\bf v}_{nm}/(\epsilon_{m\bf k} - \epsilon_{n\bf k})$ since only cross-gap matrix elements are involved, and that the 2nd $\bf A$ in each term is simply the matrix elements of the position operator $y$ in the eigenstate basis. We find that only $\delta\langle p_s \rangle_{y y}$ is nonzero, and that the gauge dependent part $\delta\langle p_s \rangle_{gd} = 0.795$ ($\hbar = e = t = E = 1$) is much larger than the gauge invariant part $\delta\langle p_s \rangle_{gi} = 5.64\times 10^{-3}$. 

\begin{figure}[h]
	\begin{center}
		\includegraphics[width=2.6 in]{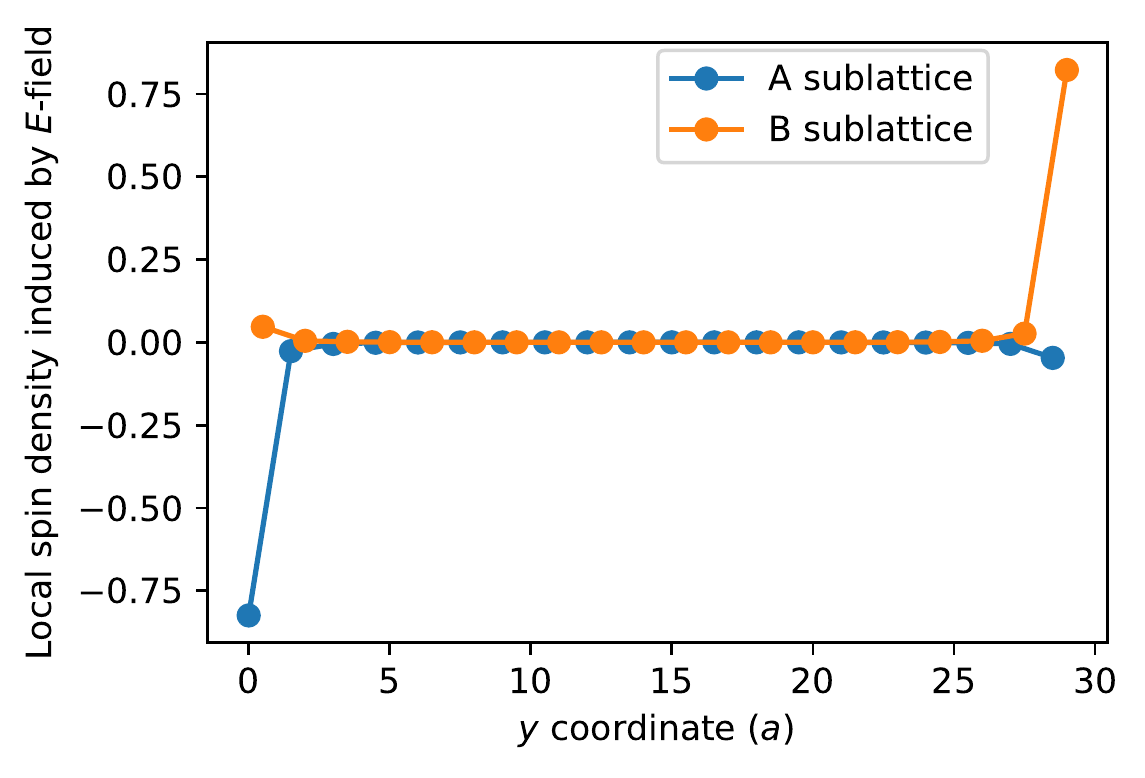}
	\end{center}
	\caption{Spatial variation of $\delta s_y$ for a zigzag ribbon with the same parameters as in Fig.~\ref{fig:KMribbon_band}.
	} \label{fig:KMribbon_lds}
\end{figure}

The nonzero value of $\delta\langle p_s \rangle_{y y}$ is due to ``accumulation" of $s_y$ near the edges of the ribbon. The local spin density induced by the electric field calculated using the quantum kinetic approach is illustrated in Fig.~\ref{fig:KMribbon_lds}. The large spin densities localized near the two edges are very close in size to $\delta\langle p_s \rangle_{gd}$, since $\delta\langle p_s \rangle_{gi}$ is negligible. For $J$ much smaller than the bulk gap this effect can be 
understood as due to balance at the edges between the $\dot{s}_z$ contribution from the bulk spin Hall effect and the torque $\tau_z =  [s_z, H ]/{\rm i}\hbar$. $s_y$ must be present at the edges to produce an expectation value for the commutator between $s_z$ and the Zeeman term, which is proportional to $s_x$.

\begin{figure}[h]
	\begin{center}
		\includegraphics[width=2.8 in]{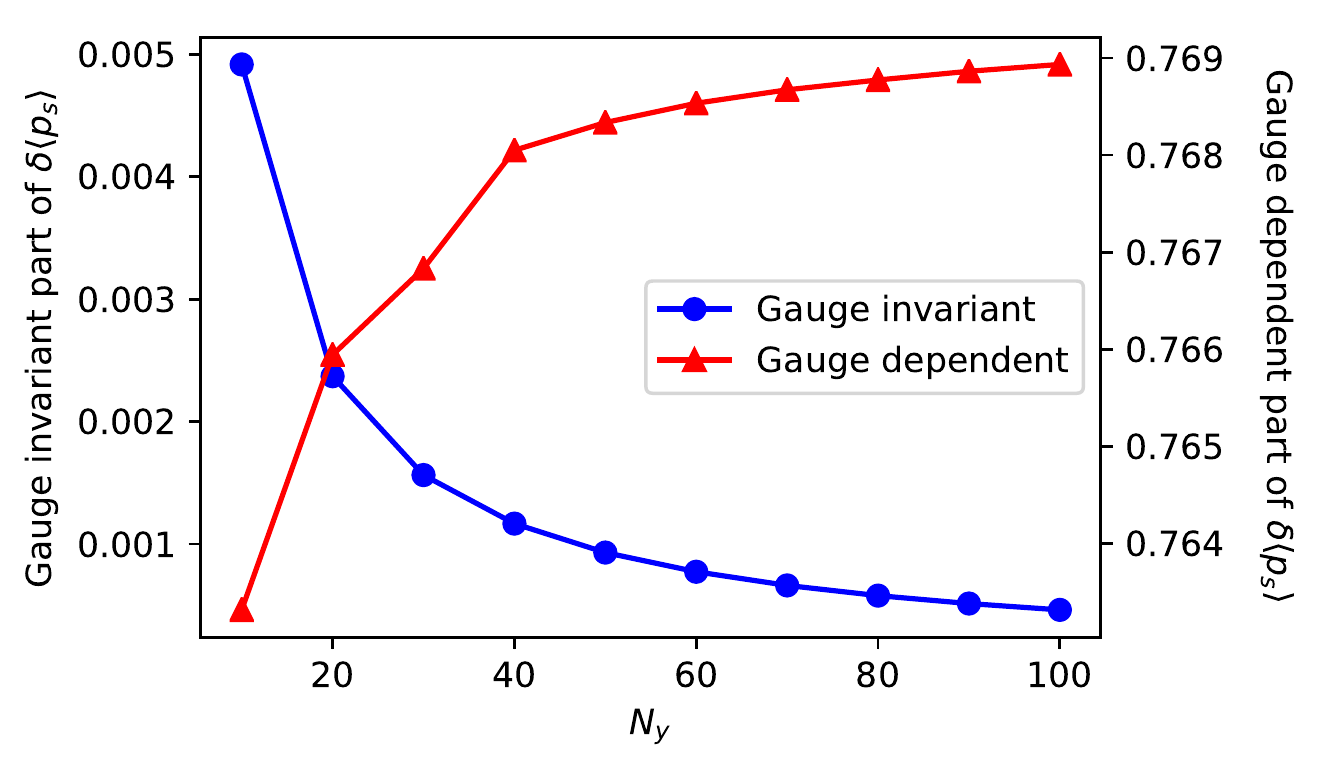}
	\end{center}
	\caption{Dependence of the gauge invariant (blue dots) and the gauge dependent parts (red triangles) of 
		$\delta \langle p_s\rangle_{yy}$ on the width of a zigzag ribbon in which
		the Zeeman field is only added to the outermost edge sites. The other parameters are the same as in Fig.~\ref{fig:KMribbon_band}.
	} \label{fig:KMribbon_dpsvsLy}
\end{figure}

As explained above, the gauge dependence of the bulk expression for $\delta\langle \overleftrightarrow{p}_s \rangle$ reflects boundary sensitivity. To illustrate this more clearly we retain the Zeeman field on the outermost edge sites of the ribbon only. The edge Zeeman fields are sufficient to open the gap and to keep the system insulating. Since the bulk of the ribbon has time-reversal symmetry, we would have $\delta\langle \overleftrightarrow{p}_s \rangle = 0 $ if there were no edges. However, Fig.~\ref{fig:KMribbon_dpsvsLy} shows that the gauge dependent part of $\delta\langle p_s \rangle_{yy}$ is almost unchanged compared to the previous case of a homogeneous Zeeman field, and saturates at a finite value as the ribbon width increases. The edge contribution can thus stay finite even in the thermodynamic limit. In contrast, the gauge invariant part decays as $1/N_y$ and vanishes in the thermodynamic limit. These behaviors are consistent with our earlier argument that the gauge dependent part can be roughly understood as the field-induced change of spin density itself, rather than as the dipole density of spin, associated with individual Wannier orbitals. We thus conclude that the SHE, even when defined in terms of strictly measurable quantities, is intrinsically sensitive to boundary conditions. 

Before ending, we point out that one can define a counterpart of the inverse spin Hall effect (ISHE) in magnetic insulators through Onsager reciprocity or thermodynamic relations. Such an ISHE is a charge polarization induced by a gradient of Zeeman field, and the reciprocal relationship is
\begin{eqnarray}
\chi_{\alpha\beta\gamma} = \tilde{\chi}_{\gamma\alpha\beta},
\end{eqnarray}
where $\chi_{\alpha\beta\gamma} = \partial(p_s)_{\alpha\beta}/\partial E_\gamma$, and $\tilde{\chi}_{\gamma\alpha\beta} = \partial p_\gamma /\partial \tilde{B}_{\alpha\beta}$, with $p_{\gamma} = er_\gamma$ the charge polarization and $\tilde{B}_{\alpha\beta} \equiv \frac{g\mu_B}{\hbar}\partial_\beta B_\alpha$ proportional to the gradient of a Zeeman field $\mathbf{B}$.

Although our discussion has been restricted to insulators, we believe the main conclusion above applies equally well to metals. Moreover, different from electric polarization, the ground state spin density polarization can be defined and discussed in metallic systems, since microscopic inhomogeneities of spin densities are not forbidden by macroscopic electromagnetism. One can, for example, introduce $\overleftrightarrow{p}_s$ as a thermodynamic variable coupled to the gradient of a Zeeman field \cite{spaldin_2013, gao_2017} in a metallic system, and attribute its linear response to electric fields to the SHE. The SHE in magnetic insulators should be observable by measuring, e.g., local Kerr response, and should generally have larger values in systems with smaller gaps, since $\delta\langle\overleftrightarrow{p}_s\rangle$ is inversely proportional to cross-gap energy differences squared (cubed for the gauge-invariant part). The gauge-invariant bulk contribution should be able to be detected with bulk probes, such as neutron or X-ray scattering, that are sensitive to inter-atomic inhomogeneities of spin densities. As for the typical size of the boundary spin accumulation coming from this SHE, if we assume that there is no order-of-magnitude difference for the inter- and intra-band matrix elements of spin and velocity operators between magnetic insulators and ordinary heavy metals, one can estimate it by comparing the gap of the insulator with $\sqrt{\hbar/\tau g(E_F)}$ of the metal, where $\tau$ is a relaxation time and $g(E_F)$ is the density of states at Fermi energy. Using the numbers given in, e.g. \cite{stamm_2017}, we find that an electric field of at least $10^3$ V/cm is needed to give an observable spin accumulation ($\sim 10^{-5}$ $\mu_B$ per atom) in magnetic insulators with gaps of $\sim 2$ eV.

\begin{acknowledgments}
HC was supported by the start-up funding from CSU. AHM was supported by SHINES, an Energy Frontier Research Center funded by the US Department of Energy, Office of Science, Basic Energy Sciences under Grant No. SC0012670, and by the Welch Foundation Research Grant No. TBF1473. QN is supported by DOE (DE-FG03-02ER45958, Division of Materials Science and Engineering), NSF (EFMA-1641101) and Welch Foundation (F-1255). The authors are grateful to Mingzhong Wu, Yasuhiro Tada, Masaki Oshikawa, Yang Gao, Di Xiao, Xin Fan, YoshiChika Otani, and Satoru Nakatsuji for helpful discussions. 
\end{acknowledgments}

\end{document}